\documentclass[%
 reprint,
nofootinbib,
 amsmath,amssymb,
 aps,
]{revtex4-2}

\usepackage{graphicx}
\usepackage{dcolumn}
\usepackage{bm}
\usepackage{upgreek}
\usepackage{subcaption}
\usepackage[export]{adjustbox}

\begin{document}

\title{Universal mask for hard X rays
}

\author{David Ceddia}
 \affiliation{School of Physics and Astronomy, Monash University, Clayton, Victoria, 3800, Australia}

\author{Alaleh Aminzadeh}%
\affiliation{Department of Materials Physics, Research School of Physics, The Australian National University, Canberra ACT 2601, Australia}

\author{Philip K.~Cook}%
\affiliation{ESRF, The European Synchrotron, 71 Avenue des Martyrs, CS 40220, 38043 Grenoble Cedex 9, France}

\author{Daniele Pelliccia}
\affiliation{Instruments and Data Tools Pty Ltd, PO Box 2114, Rowville VIC 3178, Australia}%

\author{Andrew M.~Kingston}
\affiliation{Department of Materials Physics, Research School of Physics, The Australian National University, Canberra ACT 2601, Australia, and \\  CTLab: National Laboratory for Micro Computed-Tomography, Advanced Imaging Precinct, The Australian National University, Canberra, ACT 2601, Australia}

\author{David M.~Paganin}
 \affiliation{School of Physics and Astronomy, Monash University, Clayton, Victoria, 3800, Australia~}

\date{\today}


\begin{abstract}
The penetrating power of X rays underpins important applications such as medical radiography. However, this same attribute makes it challenging to achieve flexible on-demand patterning of X-ray beams.  One possible path to this goal is ``ghost projection'', a method which may be viewed as a reversed form of classical ghost imaging. This technique employs multiple exposures, of a single illuminated non-configurable mask that is transversely displaced to a number of specified positions, to create any desired pattern.  An experimental proof-of-concept is given for this idea, using hard X rays. The written pattern is arbitrary, up to a tunable constant offset, and its spatial resolution is limited by both (i) the finest features present in the illuminated mask and (ii) inaccuracies in mask positioning and mask exposure time.  In principle, the method could be used to make a universal lithographic mask in the hard-X-ray regime.  Ghost projection might also be used as a dynamically-configurable beam-shaping element, namely the hard-X-ray equivalent of a spatial light modulator.  The underpinning principle can be applied to gamma rays, neutrons, electrons, muons, and atomic beams.  Our flexible approach to beam shaping gives a potentially useful means to manipulate such fields.   
\end{abstract}

\maketitle

\section{Introduction}\label{sec:Introduction}

The concepts established in Refs.~\cite{paganin2019writing, ceddia2022Aghost, ceddia2022Bghost} show that building signals out of noise, {\em e.g.}~building images out of random maps, is not as contradictory as it might sound. Based on these previously-published theoretical and computational studies, here we experimentally demonstrate the writing of arbitrary distributions of radiant exposure, using a single illuminated transversely-displaced non-configurable patterned random mask. In particular, we establish proof-of-concept for a universal hard-X-ray mask. The underpinning principle is very general, and can be applied to a variety of radiation and matter-wave fields---{\em e.g.}~neutrons \cite{NeutronOpticsHandbook}, electrons \cite{CowleyBook}, muons \cite{MuonRadiography, Yamamoto2020}, atomic beams \cite{AtomBeamBook}, ion beams \cite{IonBeamBook}, and gamma rays---for which configurable beam-shaping elements either do not exist, or have low spatial resolution. Three possible future applications motivate this work: (i) an X-ray spatial light modulator or data projector; (ii) a universal hard-X-ray photolithographic mask; 
(iii) 3D short-wavelength high-resolution printing in volumetric additive manufacturing \cite{Beer2019, TomographyInReverse2019, Loterie2020, Toombs2022}, as well as for sculpting desired 3D distributions of X-ray dose, {\em e.g.}~for intensity-modulated radiotherapy \cite{Cho2018}.

\begin{figure*}[ht!]
\centering\includegraphics[width=\textwidth]{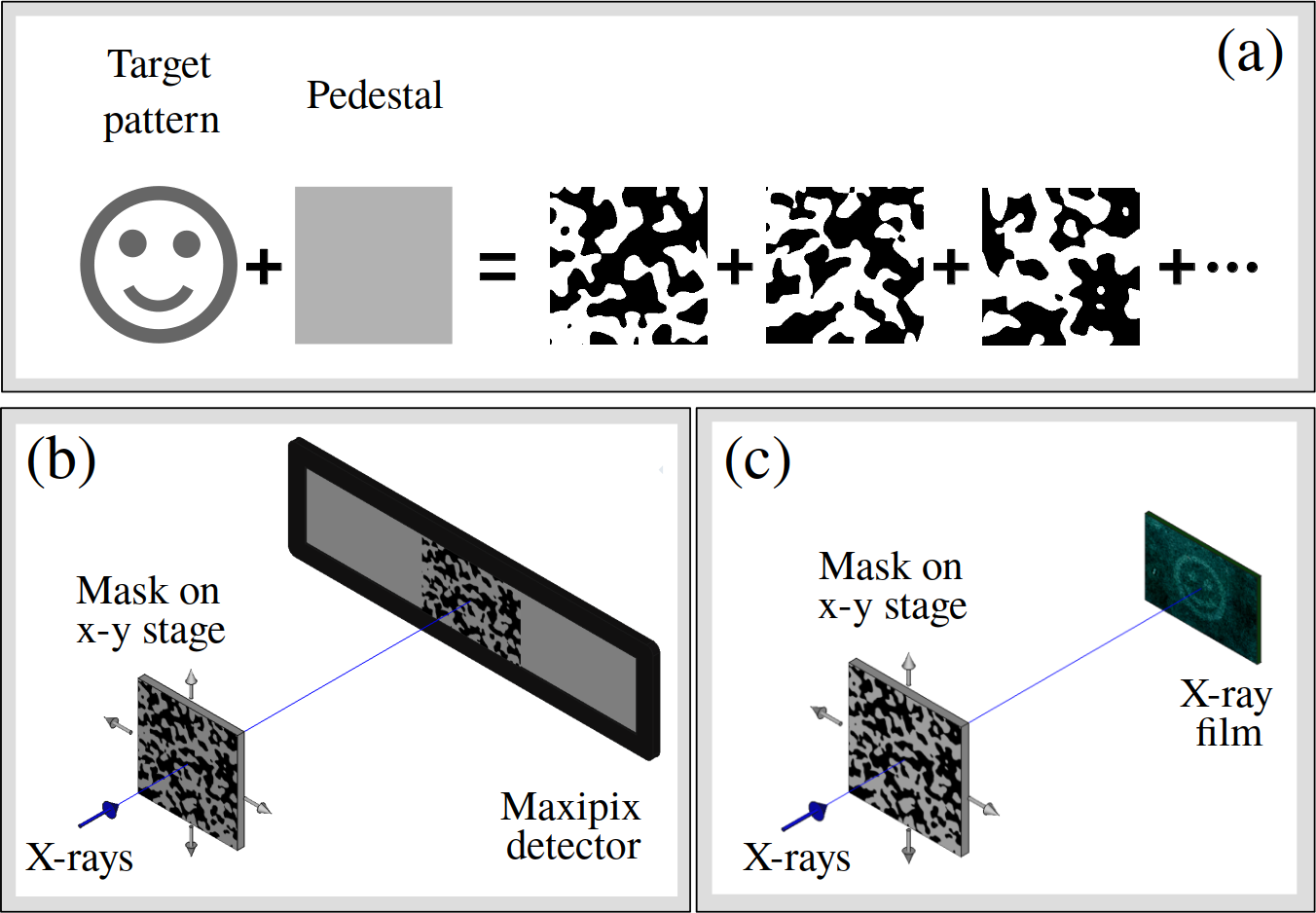}
\caption{(a) Building signals out of noise. Any 2D target pattern can be expressed as a linear combination of random speckle maps, up to an additive offset (``pedestal''). Experimental ``ghost projection'', using an illuminated random mask and (b) digital pixel-camera (Maxipix) detection or (c) film-based detection.}\label{Fig1}
\end{figure*}

The method, termed ``ghost projection'' (GP) \cite{ceddia2022Aghost, ceddia2022Bghost}, is a reversed form of classical computational ghost imaging \cite{shapiro2008computational}. Ghost imaging (GI) is an indirect imaging technique, originally developed in the context of entangled-photon quantum optics, but later shown to have a classical variant which is of primary concern here \cite{erkmen2010ghost, Shapiro2012, Padgett2017}.  GI requires the splitting of a patterned illumination (or speckled beam). One part interacts with the sample, reducing the total intensity transmitted, which is recorded by a bucket (single-pixel) detector. The second part does not interact with the sample, but rather is measured directly by a pixel array detector, forming the reference image. An image of the sample is never recorded directly. Rather, the image is reconstructed by correlating the bucket signal and reference image.  For classical illumination, the bucket measurements may be viewed as decomposition coefficients in a non-orthogonal-function expansion of the unknown sample \cite{PellicciaIUCrJ, ceddia2018random,Gureyev2018} (cf.~Refs.~\cite{Bromberg2009, Katz2009}).  Such an expansion is sketched in Fig.~\ref{Fig1}(a). 

Classical GI measures intensity correlations (bucket signals) between an unknown object and a set of patterned illuminations, in order to reconstruct the unknown object's intensity transmission function. In contrast, ghost projection---namely the method employed in the present work---seeks to {\em establish such correlations in order to create a desired image (spatial distribution of radiant exposure)}. Stated in more intuitive terms, classical ghost imaging interrogates an unknown object based on known illumination patterns and a bucket detector that cannot create a direct image, whereas ghost projection creates a desired pattern by summing together known illumination patterns using a mask that does not itself contain the desired pattern. The fundamentals of GP are developed in Refs.~\cite{paganin2019writing, ceddia2022Aghost} and an understanding of the practical considerations are explored in Ref.~\cite{ceddia2022Bghost}. Supplement 1 Section 1 provides an abstract description of GP that is based on the linear algebra of high-dimensional vector spaces, with Supplement 1 Section 2 providing a more detailed discussion comparing and contrasting ghost imaging and ghost projection.

A generic ghost-projection experiment is sketched in Figs.~\ref{Fig1}(b) and \ref{Fig1}(c). Here, a source illuminates a spatially-random non-configurable mask, thereby generating a speckle pattern over a specified illumination plane. By transversely displacing the mask, a number of different patterns may be produced. Each pattern, over the illumination plane at a distance $\Delta$ downstream of the mask, is assumed to be known. This can be because either (i) the patterns have been previously measured, or (ii) the patterns may be calculated since the structure of both the mask and the illumination has been sufficiently precisely characterized. Note, also, that while our emphasis is on random masks, non-random masks may also be employed for the purposes of ghost projection. If $\Delta$ is sufficiently large and the mask is thin, the distance between the mask and the illumination plane generates speckles via Fresnel diffraction, which may also be spoken of as propagation-based phase contrast \cite{Paganin2006} or out-of-focus contrast \cite{CowleyBook}. If $\Delta$ is sufficiently small and the mask is thin, structured illumination will instead be based primarily on mask absorption. Thick random masks may also be employed to generate speckles \cite{FinkSpeckles2012, ZhangSpecklePatterns22} for the purposes of ghost projection. Regardless of the speckle-generation scenarios, by transversely scanning the mask to specified locations, any pattern of time-integrated radiant exposure can be imprinted on the illumination plane, up to both (i) an additive constant (termed a ``pedestal,'' see Fig.~\ref{Fig1}(a)) and (ii) a limiting spatial resolution that is governed by both the finest length scale present in the speckles to a non-negligible degree \cite{paganin2019writing} and inaccuracies in mask positioning and mask exposure time \cite{ceddia2022Aghost,ceddia2022Bghost}. The key strategy of ghost projection is to {\em select a suitable set of mask translations and corresponding exposure times, such that an arbitrary distribution of integrated radiant exposure is indeed registered over the illumination plane} \cite{paganin2019writing,ceddia2022Aghost,ceddia2022Bghost}.


We close this introduction with a brief overview of the remainder of the paper. Section \ref{sec:Methods} outlines the methods that underpin our experimental demonstration of a universal hard-X-ray mask based on the ghost-projection concept, corresponding to the absorptive-mask case where $\Delta$ is sufficiently small. Section \ref{sec:Results} presents our experimental results. Some broader implications of this work are discussed in Section \ref{sec:Discussion}, followed by concluding remarks in Section \ref{sec:Conclusion}.

\section{Methods} \label{sec:Methods}

X-ray ghost projection experiments were carried out at the BM05 beamline of the European Synchrotron (ESRF) in Grenoble, France. A Si-111 double-crystal monochromator was employed with liquid-nitrogen cooling, to monochromate the beam down to a relative energy spread $\Delta E/E$ of $10^{-4}$. Two different detection modes were used for the experiment: (i) a Maxipix digital photon-counting detector with 55 $\upmu$m pixel pitch \cite{MaxipixPaper}, and (ii) X-ray film with an effective grain size of 5 $\upmu$m. Supplement 1 Section 3 contains further details on these two ghost-projection detection modes.

For the Maxipix-based experiments, our procedure consisted of raster scanning our universal mask and imaging the resulting set of patterned illuminations using the pixel detector. All mask-to-detector distances $\Delta$ were on the order of 10 mm, which is sufficiently small for the mask-illuminations patterns to be in an absorptive regime.  The raster scanning was performed using a monochromated X-ray beam of energy 23 keV and, separately, 18 keV. Electronic shuttering of the Maxipix detector, with better than 1 $\upmu$s accuracy, was used to define the exposure time. The total of $N$ images collected were then used to calculate sets of mask positions and exposure times that would generate specific target GP images; the mathematical details of our ghost-projection algorithm are given below. The horizontal mask translation stage employed had greater than 0.1 $\upmu$m repeatability, while the vertical stage had better than 3 $\upmu$m repeatability. 

Adapting these experiments to film-based detection involved (i) the installation of a physical shutter to replace the electronic shutter of the Maxipix detector, and (ii) reducing exposure times by a factor of 10 such that the dose deposition lay within the dynamic range of the film. The physical shutter was a newly-developed ESRF in-house model with open/close time of less than 30 ms \cite{FastShutter2020}.  Aluminum attenuators were used to tune the flux to an acceptable level for both the Maxipix detector and film (2.58 mm and 4.47 mm Al, respectively). 
Note that the mask was in the same position for both setups. This second set of experiments provided a clear demonstration of arbitrary intensity-patterning capability to deposit dose in a structured way on a physical object ({\em i.e.}, the X-ray film).  

The structured illumination patterns were generated using binary attenuation masks. Compared to generic random masks, binary masks are both simpler to fabricate and provide a better GP signal-to-noise ratio due to their maximum variance \cite{Kingston2021, Kingston2023}. A range of mask patterns was designed to exhibit specific properties. Specific mask designs employed here include random binary masks with different feature sizes, together with random fractal-like masks with multiscale capabilities, and a non-random Legendre mask that forms an orthogonal basis under translation.  The various types of mask were written onto a single wafer, with the ghost-projection optimization being allowed to select particular patterns from the different types of mask.  A mathematical description of each type of mask is given in Supplement 1 Section 4.

\begin{figure*}[ht!]
\centering
    \includegraphics[width=\textwidth]{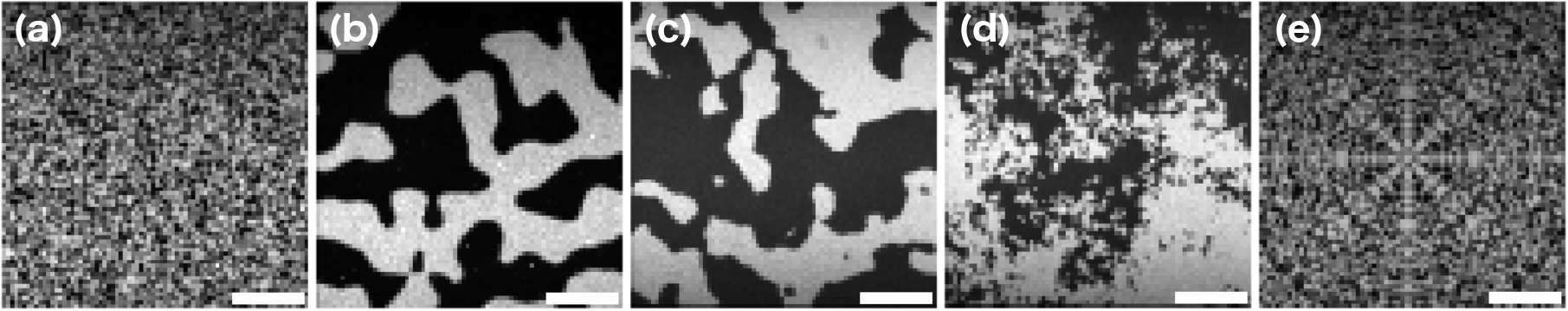}
    \caption{Universal masks employed for ghost projection using hard X rays: (a) random binary mask, (b) binarized Gaussian-smoothed-noise mask, (c) binarized Lorentzian-smoothed-noise mask, (d) random-fractal mask, (e) Legendre mask. Scale bar, in all panels, is 1 mm.}\label{fig:AlalehMasks}
\end{figure*}

Our binary attenuation masks were constructed through gold (Au) electroplating on a 4-inch diameter glass ($\textrm{SiO}_2$) substrate with a thickness of 700 $\upmu$m. The fabrication process is described in detail in Supplement 1 Section 5. The height of the electroplated mask structure was measured as 26 $\upmu$m within a 4$\%$ tolerance, as expected. At 23 keV, X-ray transmission through 26 $\upmu$m of Au is less than 8\% and the X-ray transmission is almost zero at 18 keV. See Fig.~\ref{fig:AlalehMasks} for radiographs of the five classes of fabricated binary masks employed in our GP experiments.

The procedure, to convert the mask measurements into an implementable ghost-projection scheme, began with a flux correction applied to the mask measurements to normalize the synchrotron storage ring current to 200 mA. We decided on this approach because, while normalization on the Maxipix could easily be carried out in post-processing, the film exposure requires this correction to be done in real-time. Further, a relative normalization of the masks was applied according to the maximum photon count to ensure easy interpretation and scaling of the final exposures obtained. 

Next, the two-dimensional mask images $R_{ij}$ were vectorized, mean corrected, and collated into a single matrix, $M$, as follows:
\begin{equation} \label{eq:Mask Matrix}
M = [R_{ij1} - \overline{R_{ij1}};R_{ij2}- \overline{R_{ij2}}; \cdots ;R_{ijN}- \overline{R_{ijN}}].
\end{equation}
Here the integer subscripts $(i,j)$ are pixel coordinates; the final subscript $k$ in $R_{ijk}$ denotes the $k$th image in the sequence of $N$ total images; an overline denotes the statistical average over the free indices (\textit{e.g.}, $\overline{R_{ij1}}$ is the spatial average over the pixel coordinates of the first mask).  With this notation in place, we express ghost projection as the linear algebra problem:
\begin{align} \label{eq:GhostProjection}
M \vec{w} \rightarrow \vec{I}.
\end{align}
Here $\vec{I}$ is the zero-mean, contrast-normalized, vectorized version of our target image; an overhead arrow denotes a vectorized quantity. From this point, we can determine the scheme weights $w_{k}$ via a number of methods: correlation values, correlation filtration, non-negative least squares optimization, L1-norm minimization with non-negative regularization, {\em etc}.~(as explored in Ref.~\cite{ceddia2022Aghost}). Here we used the non-negative least squares (NNLS) optimizer from \texttt{MATLAB} to solve for the weights
\begin{align} 
\text{arg~min} \|M\vec{w} - \vec{I}\|,\text{~subject to~} w_k \geq 0,
\label{eq:OptimisationProblemofGhostProjection}
\end{align}
where the latter constraint enforces that the exposures remain physical. Intuitively, this process gives a particular linear combination of the mask images that yields the desired ghost projection. The preceding equations give the essential formulae that underpin our method, with the reader being referred to Refs.~\cite{paganin2019writing, ceddia2022Aghost, ceddia2022Bghost} for additional mathematical development.   

As mentioned above, this scheme will produce a pedestal having a uniform exposure equal to $N'\overline{w_k} \overline{R_{ijk'}}$, in units of image contrast. Note that a certain pedestal can be enforced by removing the mean correction in Eq.~(\ref{eq:Mask Matrix}), and then adding the desired value (in units of image contrast) to the right-hand side of Eq.~(\ref{eq:GhostProjection}). Here $\overline{R_{ijk'}}$ is the average transmission value of the selected $N'$ masks ($k$ is used to index the ensemble of $N$ masks, and $k'$ denotes those selected for the particular target ghost-projection image). These non-negative weights are rescaled to give per-mask exposure times according to the application at hand:
we rescale the maximum expected Maxipix photon count to be within a comfortable margin of its saturation value, while for the X-ray film we scale the integrated exposure time to a predetermined value. A ghost projection is then obtained simply by recalling the locations of the $N'$ selected masks and exposing them for the predetermined period of time (with the aforementioned on-the-fly flux correction to a ring current of 200 mA).

\section{Results} \label{sec:Results}

\begin{figure*}[ht!]
    \centering
    \includegraphics[width=\textwidth]{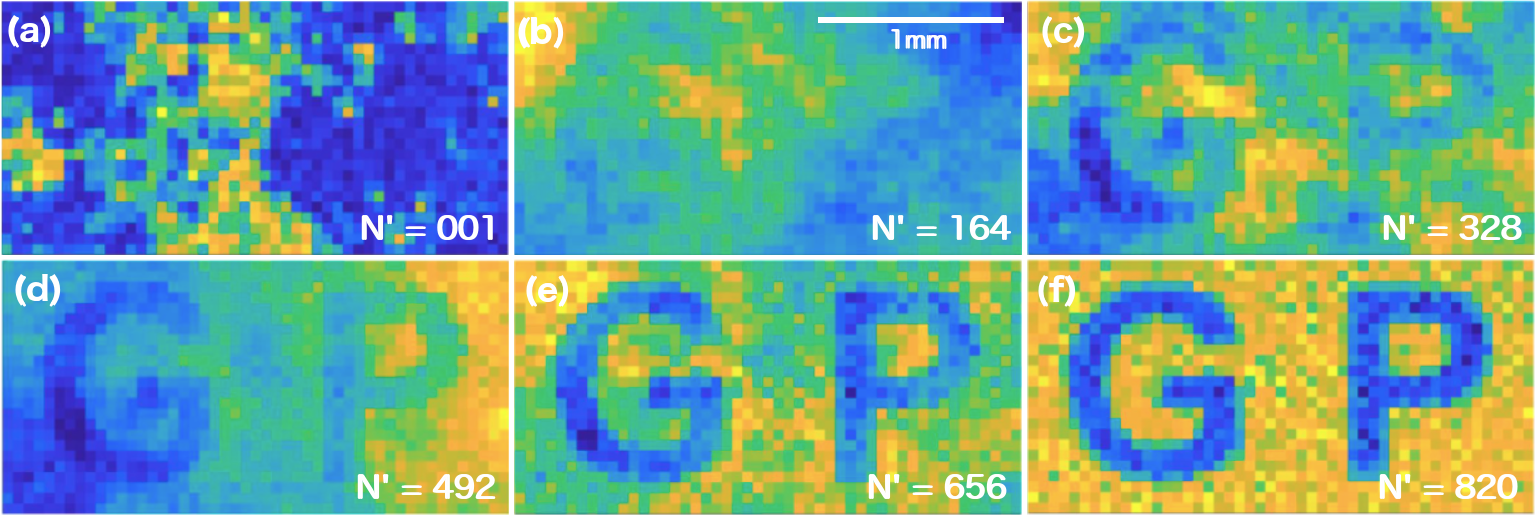}
    \caption{A subset of the sequence of cumulative ghost-projection exposures in a $25 \times 50$ pixel frame, at an X-ray energy of 23 keV, using a digital photon-counting detector.  (a) First frame in the sequence, (b) sum of the first 164 frames, (c) sum of the first 328 frames, (d) sum of the first 492 frames, (e) sum of the first 656 frames, and (f) sum of all 820 frames.  For all frames in the sequence, see the video in Visualization 1.}
        \label{fig:Four Build up GPs}
\end{figure*}

Figure~\ref{fig:Four Build up GPs} gives an experimental demonstration, of the construction of a desired image through the ghost-projection process, using 23 keV X rays.  See Visualization 1 for the corresponding video. Using transversely-displaced non-configurable mask patterns similar to Fig.~\ref{fig:Four Build up GPs}(a), with varying illumination times for each selected transverse position, over time the cumulative exposures such as those in Figs.~\ref{fig:Four Build up GPs}(b)-\ref{fig:Four Build up GPs}(e) will eventually integrate to the designed ghost projection, namely the letters ``GP'' in Fig.~\ref{fig:Four Build up GPs}(f).  A total of $N' = 820$ frames was employed in this Maxipix-based, hard-X-ray ghost projection, selected from a total pool of $N = 17\,280$ frames that were captured prior to performing the ghost projection.  A random-fractal mask was used for the first 751 frames of this ghost projection, with a binarized Lorentzian-smoothed noise-mask being used for the remaining frames. Exposure times $T$ for individual illumination frames varied between  $T_{\textrm{min}} = 1.01$ ms and $T_{\textrm{max} }= 118$ ms, with mean $\overline{T} = 20.1$ ms and standard deviation $\sigma_T = 17.3$ ms.  For further details, see Supplement 1 Section 6.

\begin{figure*}[ht!]
    \centering
    \includegraphics[width=\textwidth]{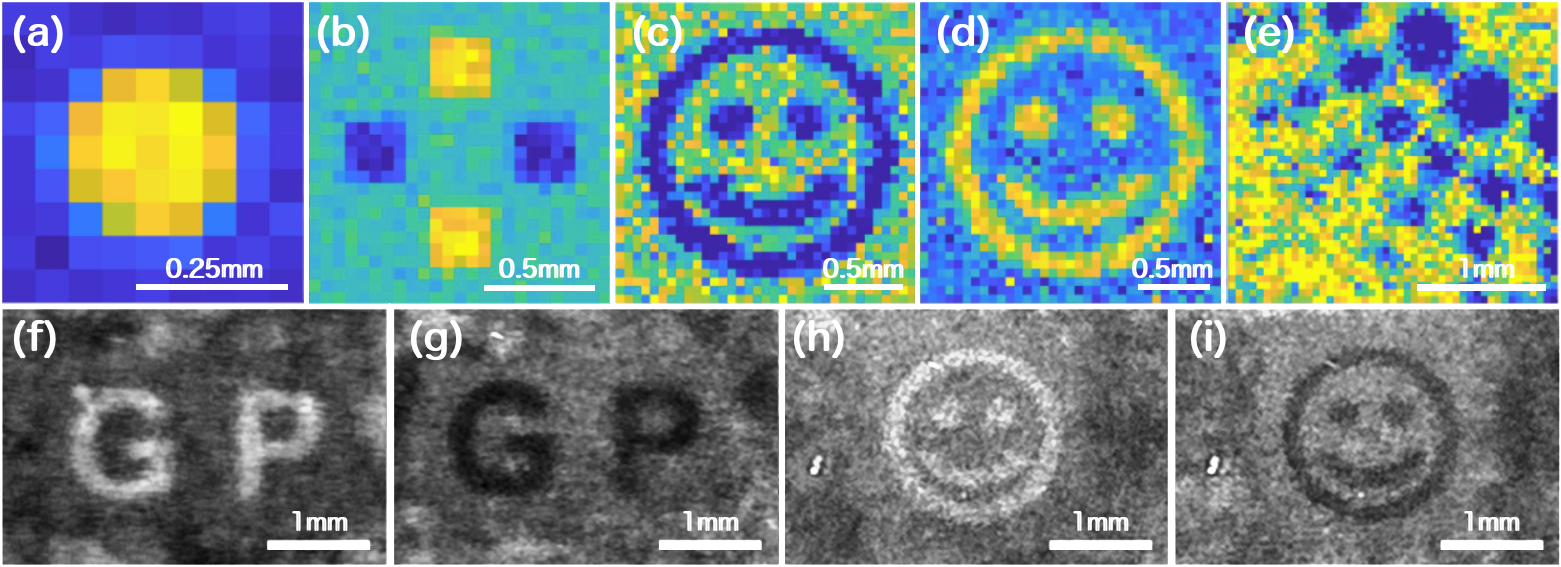}
    \caption{Example digital-detector (a-e) and film-based (f-i) X-ray ghost projections demonstrating the universality of the scheme to create various images. (a)-(b) utilize a beam energy of 23 keV, while  the remainder use a beam energy of 18 keV. }
        \label{fig:six digital GPs}
        \label{fig:Four Film GPs}
\end{figure*}

To demonstrate that ghost projection gives a universal mask for hard X rays, we projected several different distributions, from the total pool of patterns created using the same set of physical masks. We emphasize that these mask patterns inherently contain none of the desired distributions. Figures~\ref{fig:six digital GPs}(a)-\ref{fig:six digital GPs}(e) show images of increasing pixel dimensions: (a) a positive-contrast dot, (b) 2 positive-contrast and 2 negative-contrast squares, (c) a negative-contrast smiley face, (d) a positive-contrast smiley face, and (e) a negative-contrast segment of the ESRF logo.  Visualization 2, Visualization 3, Visualization 4, Visualization 5, and Visualization 6 show the full sequence of frames in the ghost projections, for each of these digital-detector cases, respectively. 

We again refer to Supplement 1 Section 6, for more detail. The number of random masks available to make each ghost projection, the selected number of random masks, the experimentally obtained signal-to-noise Ratio (SNR) and several other relevant parameters are given in Supplement 1 Table I. The  precise definition of SNR is given in Supplement 1 Section 7. A further table giving the breakdown of which mask types were employed to produce each of the digital-detector ghost projections is given in Supplement 1 Table II.  As mentioned earlier, all of the mask types were imprinted on a single wafer, with the ghost-projection optimization algorithm being used to select which masks were employed for each desired pattern of radiant exposure.


Film-based X-ray ghost projections were created using 18 keV radiation, subsequently placing the developed film on a light box and taking visible-light photographs using a Sony Alpha 7 Mark II digital camera (24.7 MP CMOS sensor). Note that contrast is reversed on X-ray film. Figures~\ref{fig:six digital GPs}(f)-\ref{fig:six digital GPs}(i) depict the successfully-produced X-ray ghost-projection film patterns: (f) a negative-contrast ``GP'', (g) a positive-contrast ``GP'', (h) a negative-contrast smiley face, and (i) a positive-contrast smiley face.

\section{Discussion} \label{sec:Discussion}

Let us now expand on the anticipated applications motivating our work, which were briefly mentioned in the opening paragraph of Section~\ref{sec:Introduction}:
\begin{enumerate}
    \item {\em Spatial-light-modulator analog, for short-wavelength radiation and matter fields:} When integrated over the time interval needed to create a desired ghost projection (cf.~Refs.~\cite{Boccolini2019, wang2020A, wang2020B}), our method may be viewed as the analog of a spatial light modulator (SLM).  It can be applied to short-wavelength radiation and matter fields---such as hard X rays, gamma rays, and neutrons---for which dynamic beam shaping elements do not currently exist with any appreciable spatial resolution. In these regimes, ghost projection has the advantage of experimental simplicity and low cost, relative to strategies which seek to construct direct SLM analogs ({\em e.g.}~with micromirror arrays \cite{Shroff2001, Chkhalo2017}).          
    \item {\em Universal mask for hard X-ray photolithography:} It is challenging to translate conventional photolithographic-mask concepts to the hard X-ray energy range, since (i) at short wavelengths the required aspect ratios for absorptive masks increase to the point where they become mechanically unstable, and (ii) the proximity effect \cite{Vladimirsky1999, Bourdillon2000, Bourdillon2001}, associated with Fresnel diffraction in the smallest available mask-to-substrate propagation distances $\Delta$, becomes stronger as feature sizes reduce.
    However, this Fresnel contrast mechanism will often be more effective than attenuation for generating high-contrast speckle patterns, 
    {\em e.g.}~in the hard-X-ray and gamma-ray domains \cite{Paganin2006}. This leads to the interesting possibility that propagation-based phase contrast \cite{Snigirev1995} might be an enabling feature of GP, in comparison to conventional methods for short-wavelength mask-based photolithography where the proximity effect 
    is typically detrimental. 
    This possibility, of ghost projection with masks whose speckle-like features exhibit contrast enhancement due to Fresnel diffraction, was examined from both a theoretical and computational perspective in Ref.~\cite{paganin2019writing}. The key idea, explored in that paper, is that a transversely-scanned random pattern---whose contrast arises wholly or in part from Fresnel diffraction---can still yield a set of linearly independent patterns for the purposes of ghost projection. In light of these comments, it would be interesting to seek to extend the absorptive-mask ghost-projection experiments of the present paper, to a propagation-based phase-contrast regime where the contrast of the speckles is improved by the influence of Fresnel diffraction. Demagnifying geometries and raster-scanning geometries might also be employed, in potential future applications of GP to short-wavelength photolithography. Non-photon lithography, {\em e.g.}~atomic-beam lithography \cite{AtomLitho1, AtomLitho2}, may also be considered from a GP perspective. In seeking to improve the spatial resolution in a lithography context, the influence of positioning accuracy will become progressively more important to mitigate, together with other factors that will limit the finest structures that can be written, such as statistical beam fluctuations.
    
    \item {\em Universal short-wavelength mask for volumetric additive manufacturing:} Volumetric additive manufacturing \cite{Beer2019,  TomographyInReverse2019, Loterie2020, Toombs2022} may be viewed as ``tomography in reverse'', where a 3D dose-sensitive substrate is illuminated from a variety of angles, to create a desired 3D distribution of radiant exposure such that a desired 3D volume is created when the exposed substrate is subsequently developed. The shorter the illumination wavelength, the finer the feature size that may be written. However, the method is currently limited by the need for high-resolution dynamic beam-shaping elements such as spatial light modulators, which do not exist in the X-ray regime. By replacing spatial light modulators with ghost projection, volumetric additive manufacturing using short-wavelength illumination (such as hard X rays) could be achieved (cf.~Refs.~\cite{paganin2019writing, KingstonOptica2018, KingstonIEEE2019}).

    \item{\em Intensity-modulated radiotherapy and microbeam radiotherapy:} GP might also be of use in shaping specified volumetric distributions of radiant exposure in the context of intensity-modulated radiotherapy \cite{Cho2018}. Microbeam radiotherapy \cite{MicrobeamRadiationTherapy2015} might also employ a ghost-projection variant, in which dose is spatially fractionated to separated high-dose volumes within a target area such as a tumor, with relatively low dose elsewhere. Synchrotron-based FLASH radiotherapy \cite{FLASH-radiotherapy-review} might also be amenable to the ghost-projection concept. 
\end{enumerate}

We now comment on the role of the additive offset, or pedestal, in the context of GP.  As shown by the example of the Hurter--Driffield curve \cite{BornWolf} for film, there is a nonlinear relation between (i) the total radiant exposure that illuminates a light-sensitive surface such as a film or photolithographic substrate, and (ii) the response of that surface or material to the illumination.  In potential future applications to photolithographic GP, the nonlinearity of the exposure--response relation could be used to advantage.  In particular, given the fact that the contrast of a ghost projection may be traded off against its associated pedestal, we can tune this additive offset \cite{ceddia2022Aghost, ceddia2022Bghost} to match the activation dose (threshold exposure) of a substrate with non-linear response. 

For the ghost projections in this experiment, the incident beam was attenuated by at least 96\% due to hardware limitations (maximum flux on the detector and maximum operating speed of the fast shutter). Even so, recording one GP required no more than a few minutes. Most of this time was in fact ``dead time'' during stage movement, so the use of GP in a production environment could benefit from high-speed stages and optimized trajectories.  Optimization of the mask-displacement trajectories, for the purposes of GP, is very closely related to the famous traveling-salesperson problem \cite{Armour1983} (see \cite{ceddia2022Bghost}). In turn, extension to an optimized shutter-free continuous-exposure GP protocol is equivalent to a continuum generalization of the traveling-salesperson problem.  

Another avenue for future work pertains to noise inclusions during the acquisition of the ensemble of masks (\textit{e.g.}~Poisson noise, ``hot''-pixels, ``dead''-pixels, beam-profile fluctuations, {\em etc}.). In our prior work, we assumed the mask acquisitions could be achieved in a mostly noise-free way and pursued non-negative least squares optimization. During our experiments, however, this proved to be a limiting factor on the final SNR obtained. A study into mask noise inclusions and a more robust GP reconstruction algorithm (such as L1-error minimization with non-negative regularization) would benefit future demonstrations and applications.

Finally, we employed a large-Fresnel-number geometry that yields a direct morphological resemblance, between the projected mask structure and the illumination pattern created by that mask. However, the GP concept is not restricted to such scenarios, as shown in the following two illustrative examples. (i) If a thick spatially-random slab is illuminated with coherent radiation or matter waves, the speckle field at the exit surface of the slab will have an intensity distribution that does not bear a direct resemblance to the morphology of the scattering volume, on account of the influence of multiple scatter \cite{FinkSpeckles2012} (dynamical diffraction \cite{CowleyBook}) within the volume of the slab. GP methods may be employed nevertheless, if the exit-surface intensity of the mask is either measured or calculable. This would enable the shaping of a desired distribution of time-integrated radiant exposure, at the exit-surface of a thick spatially-random slab, using the ghost-projection concept. 
 (ii) Consider, as a second example, an ensemble of aberrated focal fields, corresponding to a coherently illuminated circular lens whose associated collapsing spherical wave is deformed {\em e.g.}~by a phase distribution given by a suitable linear combination of Zernike circle polynomials \cite{BornWolf}.  The resulting  ensemble of aberrated focal-plane intensity distributions will be highly structured \cite{BornWolf,Paine2018} diffraction catastrophes \cite{KravtsovOrlovBook}, which could be employed as an overcomplete \cite{mandel1995optical} basis for the purposes of ghost projection. Since the diffraction physics of coherent waves in focal regions is very well understood \cite{StamnesBook}, the aberrated focal patterns need not be measured, and could instead be calculated based on the known coherent aberrations \cite{BornWolf} of the focusing system. This concept could be employed for focused-beam ghost-projection lithography using controllable-aberration coherent electron \cite{Kallepalli2022} or X-ray \cite{Kimura2013} focused probes.  Rather than transversely scanning the probe, as done {\em e.g.}~in electron-beam lithography \cite{Chen2015}, both the probe and substrate would be stationary, with an ensemble of coherent aberrations being ``dialled up'' for the probe, in order to create a desired time-integrated ghost-projection distribution of radiant exposure, over a substrate at the focal plane of the probe. 

\section{Conclusion} \label{sec:Conclusion}

An experimental demonstration was given, for use of the ghost-projection concept to create a universal mask for hard X rays.  This method, which may be viewed as a reversed form of classical computational ghost imaging, may be of relevance for applications such as a spatial light modulator for hard X rays, a universal lithographic mask for short-wavelength irradiation, volumetric additive manufacturing, and various forms of intensity-modulated radiotherapy.  While we have focused on the experimental demonstration of ghost projection using X rays, the principle might also be used for radiation and matter-wave fields for which dynamical beam-shaping elements either do not exist or have insufficient spatial resolution, such as electrons, neutrons, muons, gamma rays, atomic beams, and molecular beams. 

\section*{Funding} Australian Research Council (ARC) (DP210101312); European Synchrotron (ESRF) (proposal MI-1448).

\section*{Acknowledgments} We acknowledge the European Synchrotron (ESRF) for provision of synchrotron beam time and we thank Mathieu Manni and Nicola Vigan\`{o} for assistance and support in using beamline BM05.  We are grateful for useful discussions with Lindon Roberts, Wilfred Fullagar, and James Monro. The masks used in this work were fabricated at the Melbourne Centre for Nanofabrication (MCN) in the Victorian Node of the Australian National Fabrication Facility (ANFF).

\section*{Disclosures} The authors declare no conflicts of interest.

\section*{Data availability}
The data that support the images within this paper and other findings of this study have been stored on the mass data storage system (MDSS) at the Australian National Computational Infrastructure (NCI, nci.org.au) and are available from the corresponding author upon reasonable request. DOI for all data collected:~10.15151/ESRF-ES-955501848. 

\section*{Supplemental document} See Supplement 1 for supporting text.

\section*{Supplemental media}
See Visualization 1, Visualization 2, Visualization 3, Visualization 4, Visualization 5, and Visualization 6 for supporting videos. 

\appendix

\section*{Supplement 1, overview}

This supplement provides supporting information for the main text of our paper. Section~1 gives an abstract description of the ghost-projection concept, by linking it to the linear-algebra question of decomposing a given high-dimensional vector in terms of a superposition of random vectors. Section~2 compares classical ghost imaging and ghost projection, with the latter being viewed as a reversed or inverted form of the former. The digital-detector and film-based experimental X-ray detection modes, employed in the main text of the paper, are described in Sec.~3.  Section~4 covers the classes of ghost-projection mask design that were employed in our X-ray ghost-projection experiments.  Associated details regarding mask fabrication are given in Sec.~5. Section~6 gives additional descriptive details and experimental parameters regarding the ghost-projection images in the main text of the paper. Section~7 explains how the signal-to-noise ratio of the experimental X-ray ghost  projections was calculated.  Finally, Sec.~8 describes the ghost-projection videos that accompany this paper. 

\section*{Supplement 1, Section 1: Abstract description of ghost projection}

Consider a large number of random vectors having no preferred direction, the tails of which are all fixed to a given point that is specified to be the origin of coordinates.  To approximate any desired vector as a linear combination of these random vectors, we can  (i) discard every random vector that has a negative projection with respect to the desired vector, before (ii) subsequently summing the vectors that remain, and then (iii) multiplying by a suitable positive constant.  More efficient vector-selection schemes could of course be chosen, but the key in-principle concept is clear, namely that random vectors form a mathematical basis \cite{Gorban2016}. This geometrically-framed idea works in any number of dimensions, so we can consider our vector space to be a high-dimensional function space. Each vector, in this high-dimensional space, may be associated with a distinct image \cite{BarrettMyersBook}. Suppose, now, that each axis of this function space corresponds to a linearly-independent two-dimensional noise map (``speckle field''), such as might be produced by transversely scanning an illuminated spatially-random screen. We immediately conclude that an arbitrary image---or, stated more precisely, an arbitrary time-integrated spatial distribution of radiant exposure---may be synthesized by transversely scanning a spatially-random illuminated screen \cite{paganin2019writing, ceddia2022Aghost, ceddia2022Bghost}.  The arbitrariness of this ``ghost projection'' image is up to a spatial resolution dictated by (i) the finest features present in the patterned illumination \cite{paganin2019writing}, as well as (ii) inaccuracies in mask positioning and mask exposure time \cite{ceddia2022Aghost, ceddia2022Bghost}.  Also, there will be an additive offset (``pedestal'') in the projected pattern, since intensity measurements can never be negative.  The previously-mentioned pedestal can be reduced, to a limited extent, via a tradeoff that balances reduced pedestal against increased noise in the ghost-projection signal \cite{ceddia2022Aghost,ceddia2022Bghost}. See Fig.~1(a) of the main paper, for a schematic indication of these enabling concepts.

\section*{Supplement 1, Section 2: Comparison of ghost projection with ghost imaging}

The notion of ghost projection \cite{paganin2019writing,ceddia2022Aghost,ceddia2022Bghost} may be viewed as a reversed form of computational classical ghost imaging \cite{shapiro2008computational, Shapiro2012, Padgett2017}.  In the latter technique, the intensity--intensity correlations (``bucket'' signals) between predetermined illumination masks and an unknown illuminated object are employed, in order to calculate an image of that object.  In the former technique, intensity correlations are established over the ghost-projection plane, rather than being measured via a bucket detector. Both positive and negative correlations may be established, as we have already seen in our ability to generate patterns with both positive and negative contrast relative to the ghost-projection pedestal (see Fig.~4(b) in the main paper).  

While ghost projection may be viewed as a reversed form of classical computational ghost imaging, the role of prior knowledge is very different.  (i) For ghost imaging with random masks in the absence of any prior knowledge regarding the sample, a relatively large number of bucket measurements needs to be taken, on account of the random-mask basis being non-optimal in comparison to complete orthogonal-basis mask sets.  In the presence of suitable prior knowledge regarding the sample, the number of required bucket measurements may be reduced (in classical computational ghost imaging), thereby decreasing the data-acquisition time and reducing the dose to the sample.  (ii) For ghost projection, however, we necessarily have total prior knowledge regarding the particular image that we wish to ghost project. This enables us to select a very small fraction of our masks, in order to generate a ghost projection using relatively few masks.  Importantly, if two or more independently-translatable random masks are illuminated in series \cite{KSMacknowledgment}, the number of possible random masks in the resulting overcomplete \cite{mandel1995optical} basis is exponentially large \cite{ceddia2022Aghost, ceddia2022Bghost}.  This allows us to choose an extremely small fraction of the possible random masks, enabling an efficient ghost projection with far fewer masks than would be needed if we worked with a particular specified set of orthogonal masks which all needed to be used.  

Returning to the ``sheaf of random vectors'' \cite{paganin2019writing, ceddia2022Aghost} concept in Sec.~1 of this Supplement, the ability to discard most random vectors evidently allows us to work with a chosen subset that enables particularly efficient ghost projection (cf.~Ref.~\cite{Nakanishi}). Comparisons with compressed sensing are natural, in this context \cite{CandesTao2006, Donoho2006,  Rani2017}.  

Ghost projection has total prior knowledge of the image to be projected, a situation having no direct analog in ghost imaging. This complete prior knowledge implies that the question of optimal mask choice is different for computational ghost imaging and ghost projection.  A variety of masks can be employed in either case, such as random binary masks, random fractal masks, uniformly redundant arrays, and masks based on the finite Radon transform~\cite{Kingston2023}. It would be interesting to investigate which class of mask is most appropriate for particular ghost-projection applications, in future explorations of the method.  In a similar vein, while we have employed non-negative least squares as the means by which the illuminated subset of possible masks is chosen, the use of more sophisticated computational optimization approaches will likely make the process more efficient.

This leads to the closely-related question of the limiting contrast achievable using ghost projection.   (i) For the inefficient fully-analytical ghost-projection scheme proposed in Ref.~\cite{paganin2019writing}, approximate formulae were developed for the maximum attainable contrast (see Eqs.~(86) and (91) therein).  This gives a pessimistic answer, on the order of 14\%, for the particular ghost-projection scheme presented there. However, it should be noted that Ref.~\cite{paganin2019writing} considered only positive correlations in the context of ghost projection, whereas Fig.~4(b) from the main paper shows that negative correlations may also be established.  Including both negative and positive ghost-projection contrast enables us to double the previous pessimistic answer, to give a limiting contrast on the order of 28\%.  (ii) For the more efficient algorithm developed in Refs.~\cite{ceddia2022Aghost,ceddia2022Bghost}, the computer modeling therein reported gives a maximum ghost-projection contrast on the order of 30\%. This is broadly consistent with the experimental findings of the present paper. It is currently unclear to what extent this contrast can be improved, from a practical perspective. Broadly speaking, the practical limits on the ghost-projection contrast will likely be related to the size of the mask, the design of the mask, the mask contrast and the nature of the final images that one wishes to ghost-project. (iii) Interestingly, in the absurdly idealized limit of an infinitely large spatially-random mask, the fundamental limit on the ghost-projection contrast is 100\% because---in principle, but not in practice---every desired ghost-projection pattern will be contained somewhere within an infinitely-large random mask.

\section*{Supplement 1, Section 3: Experimental detection modes}

Two different detection modes were used for the X-ray ghost-projection experiments: 
\begin{description}
    \item [Maxipix] A digital photon-counting pixel detector developed at ESRF, Maxipix (Multichip Area X-ray detector based on a photon-counting PIXel array) \cite{MaxipixPaper}. The Maxipix has a $1260 \times 256$  pixel array with a pixel pitch of 55 $\upmu$m. 
    \item [X-ray film]  Structurix D3-SC industrial X-ray film (Agfa-Gevaert Group; Mortsel, Belgium). The film was processed using an Industrex M37 plus Processor (Colenta Labortechnik GmbH; Wiener Neustadt, Austria) with XR D-6 NDT developer and XR F-6 NDT fixer solutions (Duerr NDT; Bietigheim-Bissingen, Germany). The film has a 3 $\upmu$m physical grain size.  The effective grain size is 5 $\upmu$m, considering errors from exposure, developing, and digitization.
\end{description}

\section*{Supplement 1, Section 4: Mask design}

The following classes of mask were employed for our ghost-projection experiments, with further details available in Ref.~\cite{Kingston2023} in a classical-ghost-imaging context.  Note that, in the experiments reported in the main paper, all of the following types of mask were written on a single wafer.
\begin{itemize}
\item {\em Random binary masks} $\mathcal{A}$ correspond to binary random noise on a pixelated array of contiguous square plaquettes, with each of two different transmission values at each pixel being equally likely.  The random transmission coefficient for each pixel is statistically independent of every other pixel, by construction. 
\item {\em Binarized Gaussian-smoothed noise masks} $\mathcal{B}_{\sigma}$ correspond to convolving $\mathcal{A}$ with a Gaussian function of specified full width at half maximum (FWHM), equal to $2\sqrt{2\ln{2}}\sigma$ pixels, and then binarizing the resulting array.
\item {\em Binarized Lorentzian-smoothed noise masks} $\mathcal{C}_{\gamma}$ correspond to convolving $\mathcal{A}$ with a Lorentzian function of specified FWHM, equal to 2$\gamma$ pixels, and then binarizing the resulting array.
\item {\em Random-fractal masks} $\mathcal{D}_{\alpha,\beta}$ correspond to convolving $\mathcal{A}$ with a suitable filter kernel, followed by binarization.  The filter kernel, for this approximately self-similar mask \cite{SethnaBook}, has the Fourier-space form
\begin{equation}\label{eq:fractal}
    H(k_x,k_y) = \frac{1}{\left(k_x^2 + k_y^2\right)^{\alpha/2} + \beta}.
\end{equation}
Here, $(k_x,k_y)$ denote discrete spatial frequencies in each of two transverse dimensions \cite{Press2007}, $\alpha \ge 0$ is a critical exponent \cite{Huang_1987} that governs the power-law decay of the fractal-mask power spectrum at large spatial frequencies, and $\beta$ is a small real regularization parameter which mollifies the blowup that would otherwise occur at the Fourier-space origin $(k_x,k_y)=(0,0)$. The special case $\alpha = 2$ corresponds to a Lorentzian function.
\item {\em Legendre masks} $\mathcal{E}_{p}$ corresponds to a 2D $p \times p$ pattern, constructed from the finite Radon transform, that is orthogonal under translation.  This  mask is defined in Ref.~\cite{petersen2022curious}.
\end{itemize}

The classes of universal mask employed for ghost projection using hard X-rays (as depicted in Fig.~2 in the main text of the paper) are  as follows: (a) random binary mask $\mathcal{A}$, (b) binarized Gaussian-smoothed-noise mask $\mathcal{B}_{\sigma}$ ($\sigma =$~8.5 pixels), (c) binarized Lorentzian-smoothed-noise mask $\mathcal{C}_{\gamma}$ ($\gamma =$~14.14 pixels), (d) random-fractal mask $\mathcal{D}_{\alpha,\beta}$ ($\alpha =$~1, $\beta$~=~0), (e) Legendre mask $\mathcal{E}_{p}$, $p = 127$.

\section*{Supplement 1, Section 5: Mask fabrication}

The attenuation ghost-projection masks were fabricated on a 4-inch $\textrm{SiO}_2$ substrate (wafer) with 700 $\upmu$m thickness. The fabrication process was mainly carried out at the Melbourne Centre for Nanofabrication (MCN) in Melbourne, Australia. The substrate was first cleaned with piranha solution to remove organic residues and then coated with 20 nm of Cr followed by 100 nm of Au using an Intlvac Nanochrome sputtering machine. The Cr layer was used as an adhesion layer because the adhesion between $\textrm{SiO}_2$ and Au is weak. The Au layer was used as a conductive layer for the subsequent electroplating. After the sputtering process, the wafer was spin-coated with an AZ 40XT-11D positive photoresist and baked at 126 $^{\circ}$C for 5 min. The thickness of the photoresist was measured as approximately 40 $\upmu$m using an optical profilometer (Bruker Contour GT-I). This thickness is sufficient to obtain 20 to 30 $\upmu$m electroplated structures. Our mask patterns were then transferred from a Cr mask into the photoresist under UV light exposure and through a photolithography process (using the EVG6200 mask aligner instrument). The exposed wafer was baked at 105 $^{\circ}$C for 90 s before it was developed in an AZ 726 solution for 5 min. In the last step of the fabrication process, Au electroplating was performed in a Pur-A-Gold 402 solution (trademark Macdermid-Enthone) for 90 min. The height of the electroplated mask structures was measured as 26 $\upmu$m within a 4$\%$ tolerance, as expected.

\section*{Supplement 1, Section 6: Details of example ghost-projection images}

The cumulative exposures in the first experimental X-ray ghost projection, as presented in Fig.~3 of the main text of the paper, are reproduced along with additional numerical parameters (scale bars indicating photon counts, and additional detail in the caption) in Fig.~\ref{fig:Four Build up GPs==SUPP} below.

\begin{figure*}[ht!]
     \centering
     \begin{subfigure}{0.33\textwidth}
         \centering
         \includegraphics[width=\textwidth]{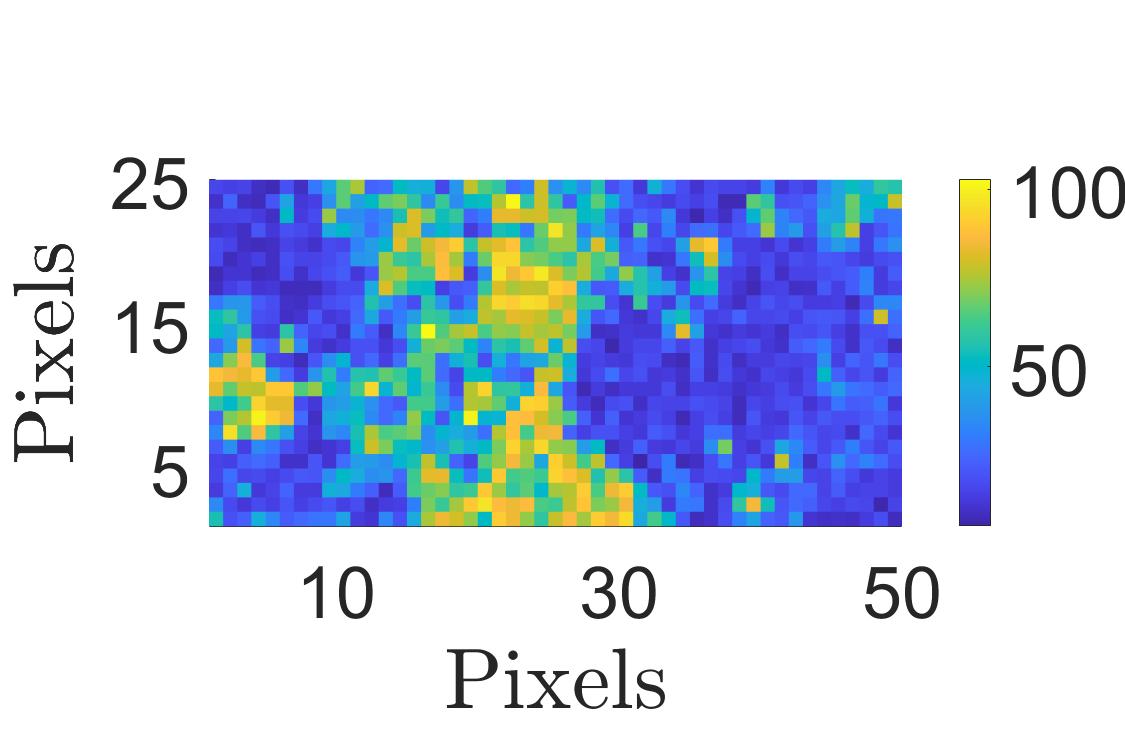}
         \caption{$N' = 1/820$}
         \label{subfig: GP build up 1}
     \end{subfigure}
     \begin{subfigure}{0.32\textwidth}
         \centering
         \includegraphics[width=\textwidth]{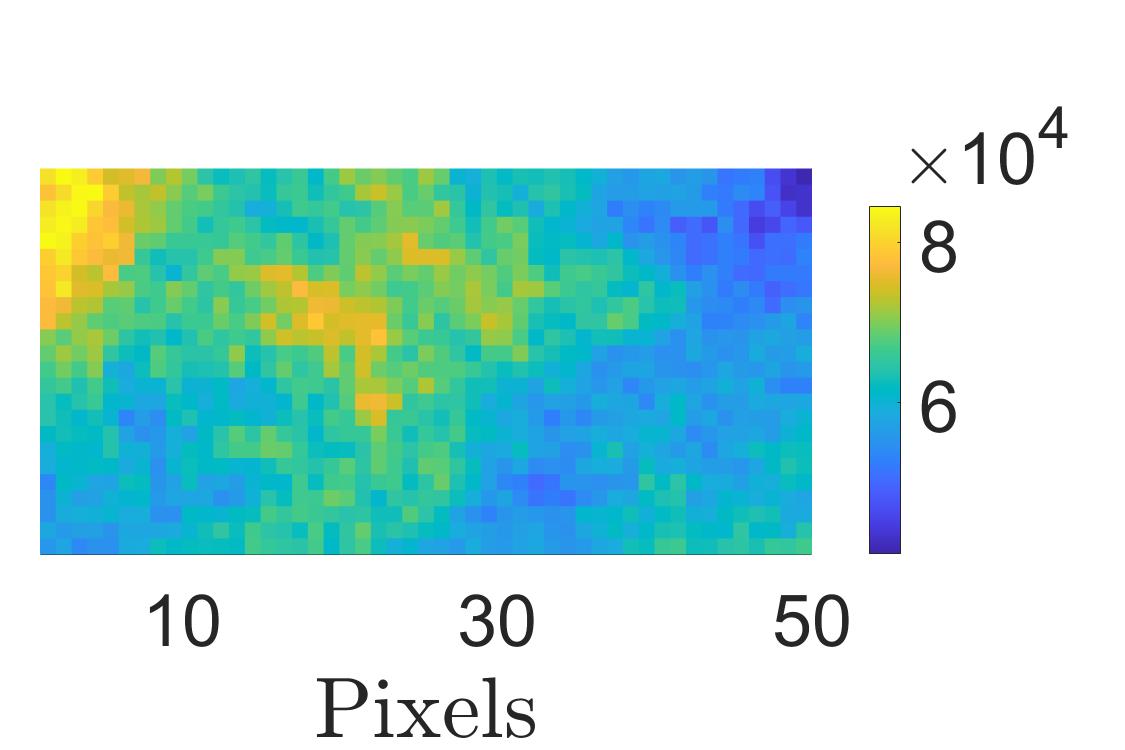}
         \caption{$N' = 164/820$ }
         \label{subfig: GP build up 2}
     \end{subfigure}
     \begin{subfigure}{0.32\textwidth}
         \centering
         \includegraphics[width=\textwidth]{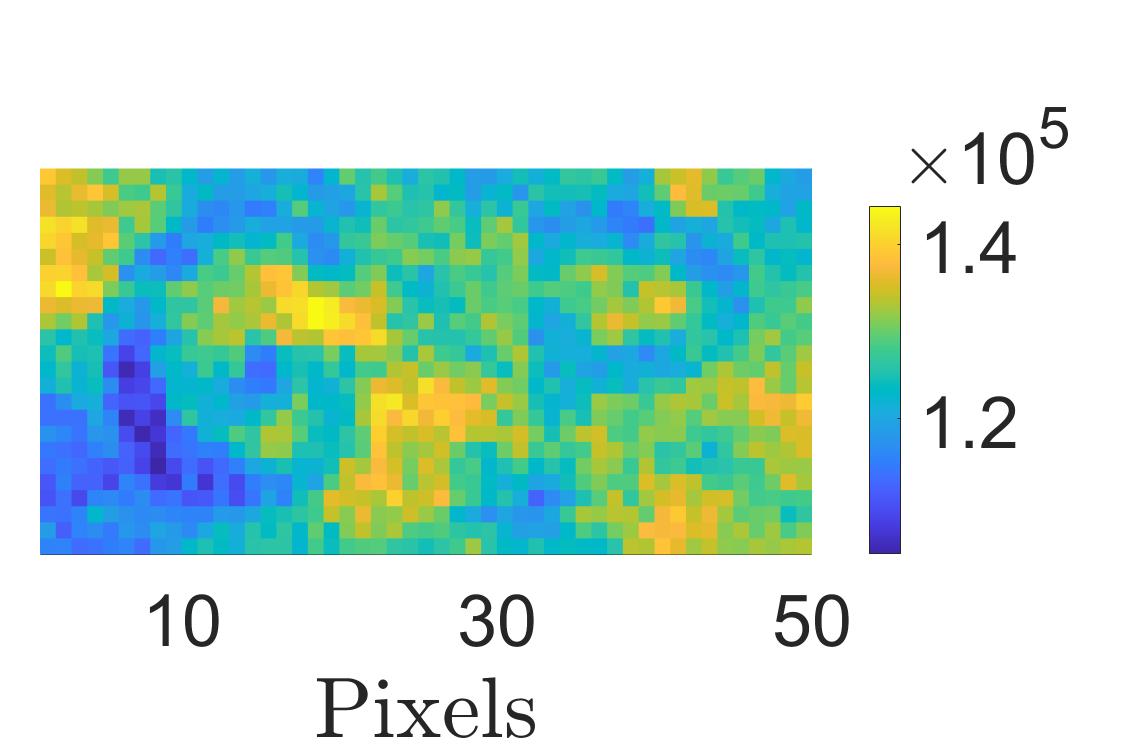}
         \caption{$N' = 328/820$ }
         \label{subfig: GP build up 3}
     \end{subfigure}
     \begin{subfigure}{0.33\textwidth}
         \centering
         \includegraphics[width=\textwidth]{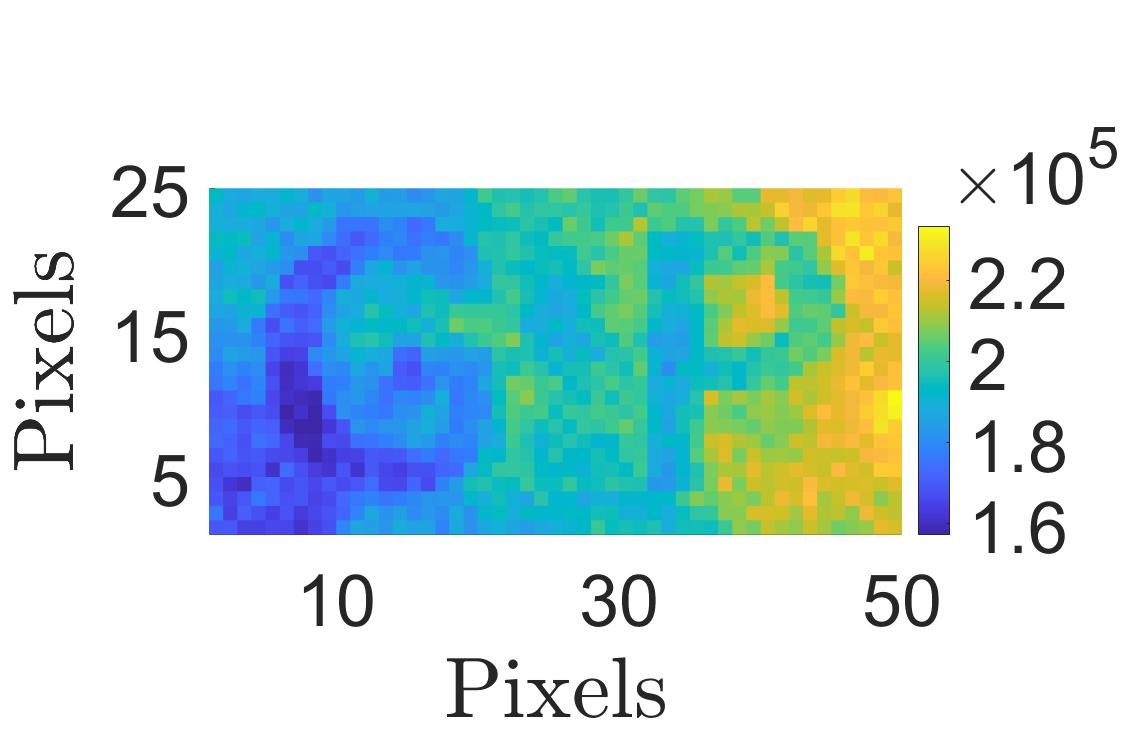}
         \caption{$N' = 492/820$ }
         \label{subfig: GP build up 4}
     \end{subfigure}
     \begin{subfigure}{0.32\textwidth}
         \centering
         \includegraphics[width=\textwidth]{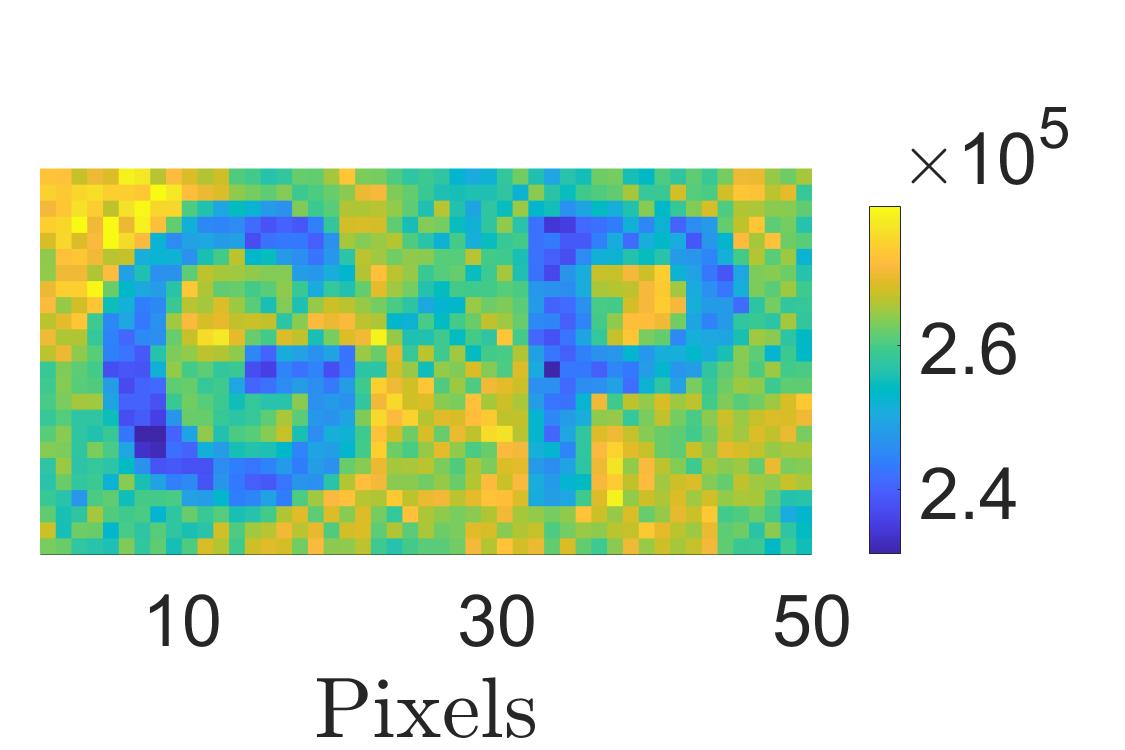}
         \caption{$N' = 656/820$ }
         \label{subfig: GP build up 5}
     \end{subfigure}
     \begin{subfigure}{0.32\textwidth}
         \centering
         \includegraphics[width=\textwidth]{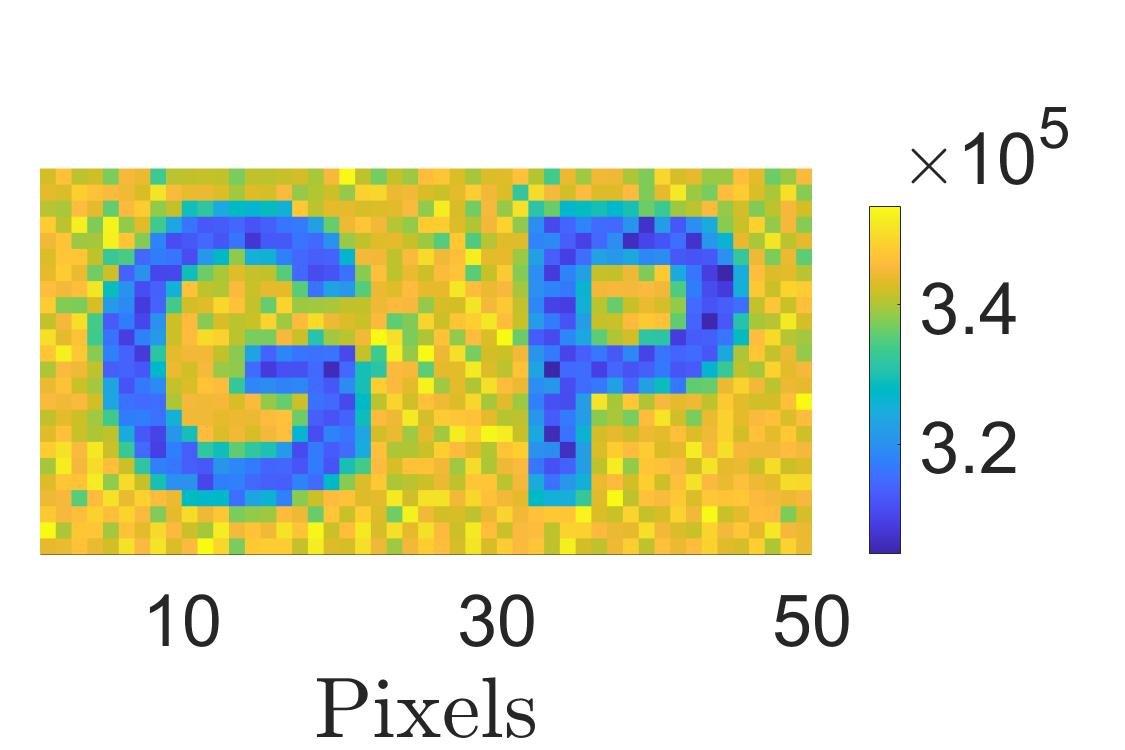}
         \caption{$N' = 820/820$ }
         \label{subfig: GP build up 6}
     \end{subfigure}
        \caption{Sequence of cumulative 23 keV X-ray ghost-projection exposures in a $25 \times 50$ pixel frame (units of photon counts). Maxipix pixel pitch is 55~$\upmu$m.  The video in Visualization 1 shows the full sequence of $N'=820$ frames that were selected from a total pool of $N=17\,280$ frames, for both (i) the $N'$th random-mask image in the exposure and (ii) the cumulative exposure for the first $N'$ masks.}
        \label{fig:Four Build up GPs==SUPP}
\end{figure*}

We now give additional information regarding the X-ray ghost projections in Fig.~4 from the main text of the paper.  This additional detail is given in Fig.~\ref{fig:six digital GPs==SUPP} below. From a given set of random masks, we projected, in order of increasing total pixels:
\begin{itemize}
\item a 10$\times$10 pixel image of a 
dot,
using the Maxipix detector (Fig.~\ref{subfig: GP1}); 
\item a 26$\times$26 pixel image containing two ``up'' dots and two ``down'' dots, where ``up'' and ``down'' refer to regions of relatively high and low dose, respectively, using the Maxipix detector (Fig.~\ref{subfig: GP2}); 
\item a 25$\times$50 pixel image of the initials ``GP'', defined by a region of relatively low dose, using the Maxipix detector (Fig.~\ref{subfig: GP3}) and film (Fig.~4(f) in the main paper);
%
\item an ``inverse GP'' image on film, Fig.~4(g)
in the main paper, where by ``inverse'', we mean that the regions of relatively high exposure are inverted to relatively low, and vice versa; 
\item a 36$\times$36 pixel image of a smiley face defined by a region of relatively low dose, using the Maxipix detector (Fig.~\ref{subfig: GP4}), and on film (Fig.~4(h) in the main paper);

\item an inverse smiley distribution on the Maxipix detector (Fig.~\ref{subfig: GP5}) and on film (Fig.~4(i) in the main paper);
%
\item a 44$\times$48 pixel image of one-quarter of the ESRF's logo dots (Fig.~\ref{subfig: GP6}). 
\end{itemize}

\begin{figure*}[ht!]
     \centering
     \begin{subfigure}{0.329\textwidth}
         \centering
         \includegraphics[width=\textwidth]{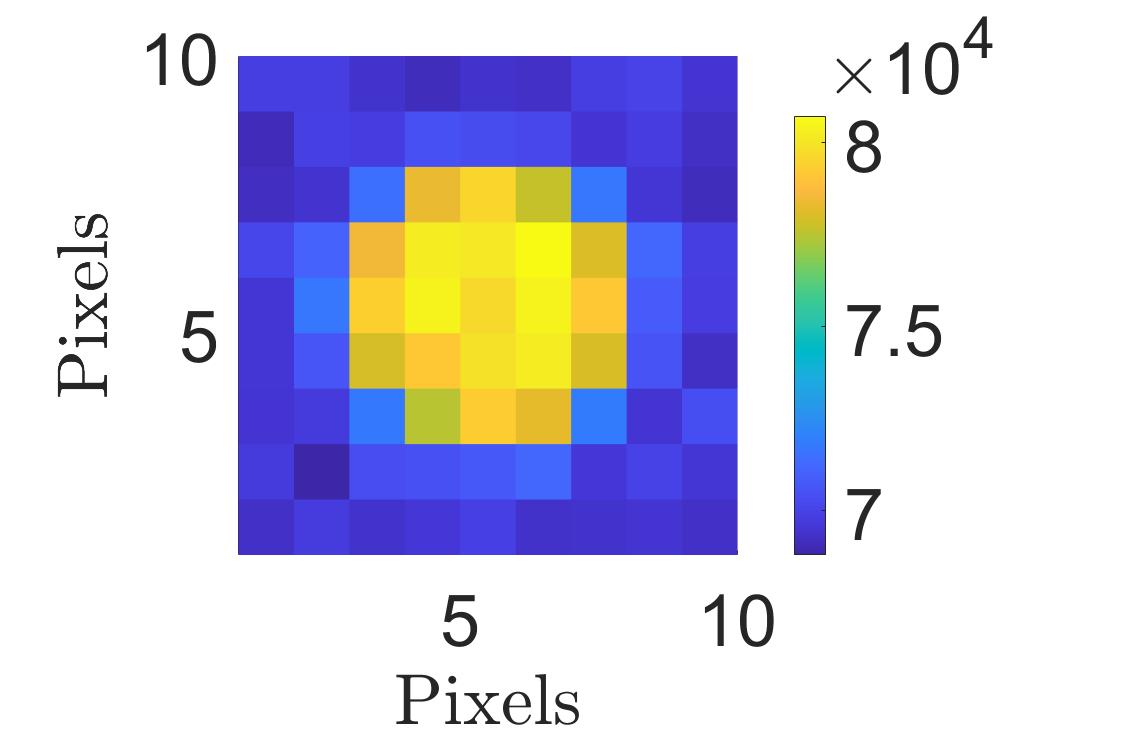}
         \caption{ }
         \label{subfig: GP1}
     \end{subfigure}
     \begin{subfigure}{0.329\textwidth}
         \centering
         \includegraphics[width=\textwidth]{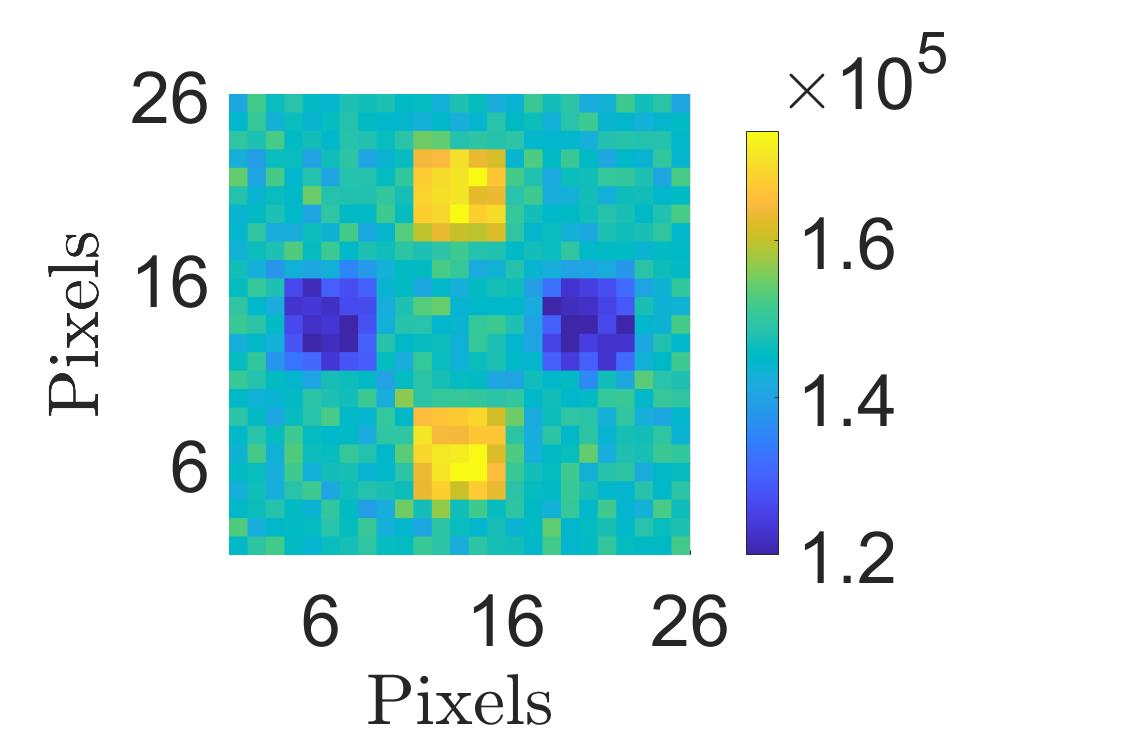}
         \caption{ }
         \label{subfig: GP2}
     \end{subfigure}
     \begin{subfigure}{0.329\textwidth}
         \centering
         \includegraphics[width=\textwidth]{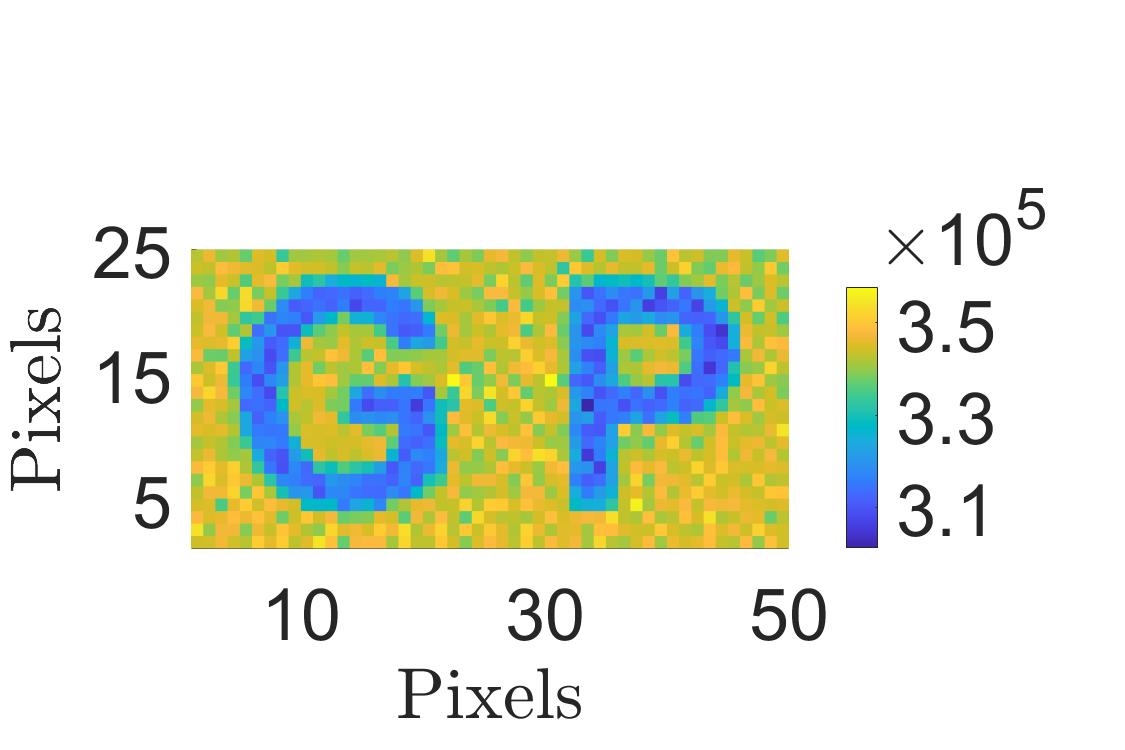}
         \caption{ }
         \label{subfig: GP3}
     \end{subfigure}
     \begin{subfigure}{0.329\textwidth}
         \centering
         \includegraphics[width=\textwidth]{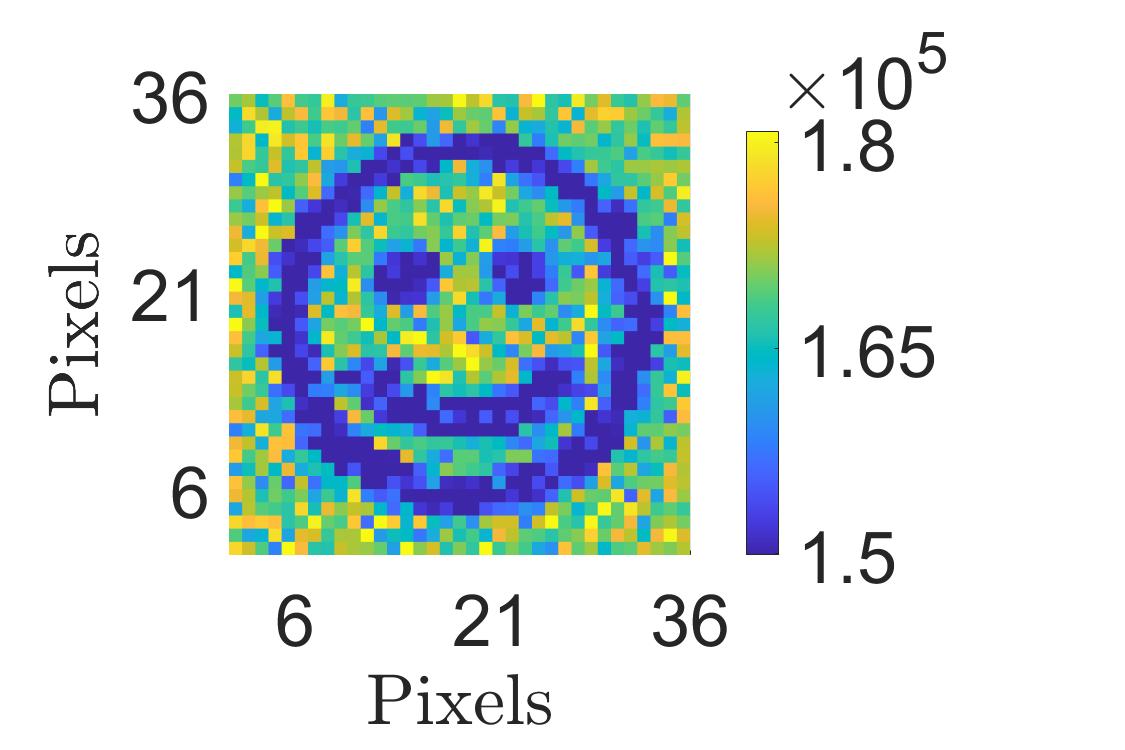}
         \caption{ }
         \label{subfig: GP4}
     \end{subfigure}
     \begin{subfigure}{0.329\textwidth}
         \centering
         \includegraphics[width=\textwidth]{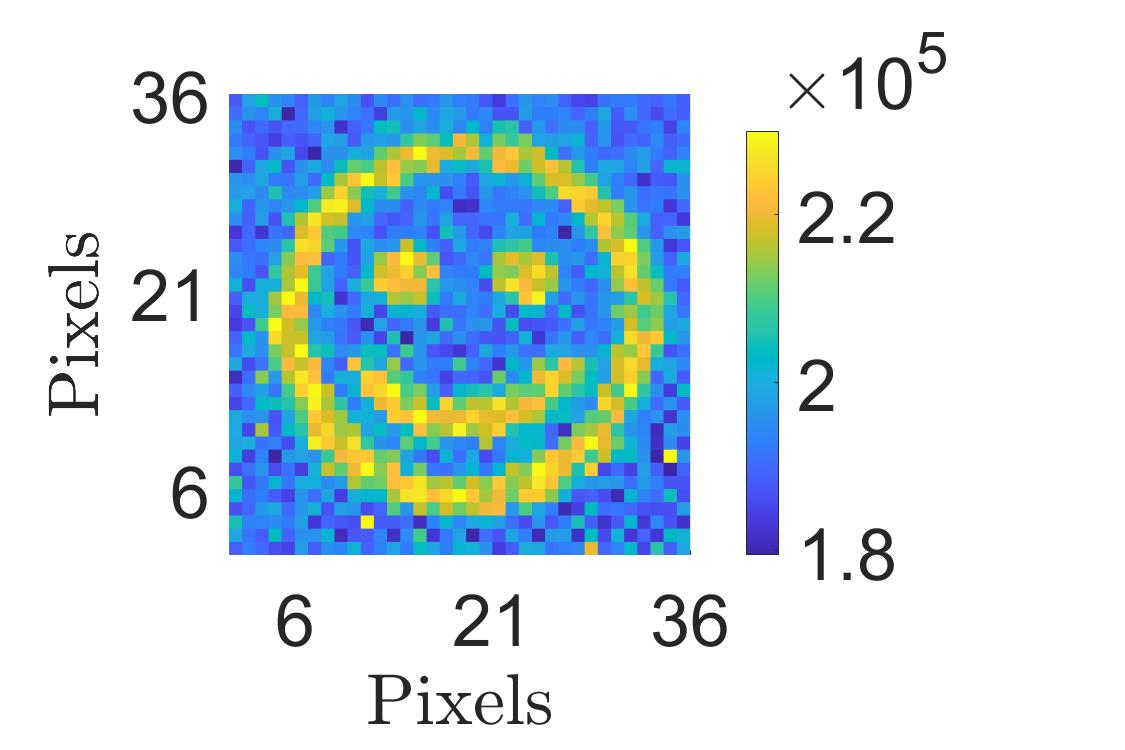}
         \caption{ }
         \label{subfig: GP5}
     \end{subfigure}
     \begin{subfigure}{0.329\textwidth}
         \centering
         \includegraphics[width=\textwidth]{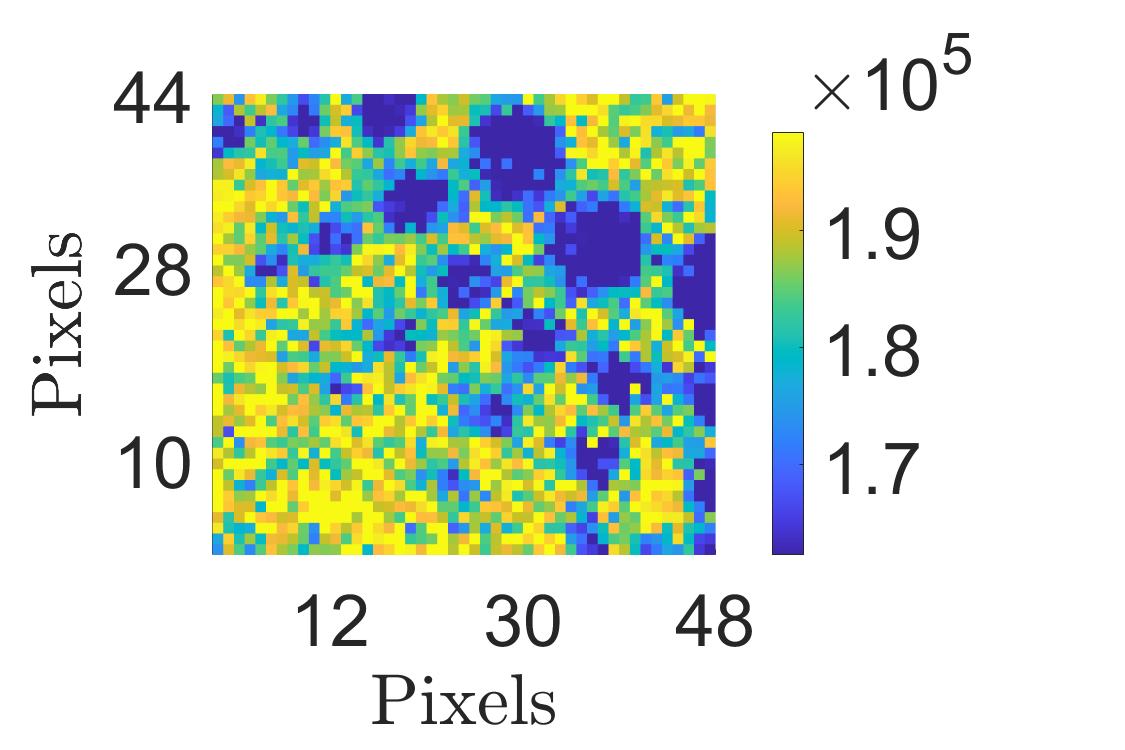}
         \caption{ }
         \label{subfig: GP6}
     \end{subfigure}
        \caption{Maxipix X-ray ghost projections, expressed in photon counts, demonstrating the universality of the scheme to create varied projections, in order of increasing number of pixels. (a)-(c) are X-ray ghost projections with beam energy 23 keV, (d)-(f) use a beam energy of 18 keV. 
        Associated videos show the full sequence of frames in the ghost projections for the respective panels: (a) Visualization 2, (b) Visualization 3, (d) Visualization 4, (e) Visualization 5, and (f) Visualization 6.  Each video shows (i) each random-mask image in the exposure and (ii) the cumulative exposure. Additional relevant parameters are given in Table~\ref{table:Mask/Image GP} and Table~\ref{table:WhichMasksWereUsedForWhichGPs}.      
        }
        \label{fig:six digital GPs==SUPP}
\end{figure*}

\section*{Supplement 1, Section 7: Determination of signal-to-noise ratio}

The signal-to-noise ratio (SNR) of each Maxipix-based GP result was determined by first calculating the variance. In particular, the variance of the experimental ghost projection $P$ was estimated by subtracting the average (pedestal, $\overline{P}$), {\em i.e.}, $P' = P - \overline{P}$, and rescaling the result to have a consistent standard deviation with the target image, $I$. The target image was then subtracted, the result squared, and the pixels were summed over in the usual way:
\begin{align}
    \text{Var}[P'] =  \text{E} \left[ \left( P' \sqrt{\frac{\text{E}[I^2]}{\text{E}[P'^2]}} - I 
 \right)^2 \right].
\end{align}
Above, ``E'' denotes expectation value and ``Var'' denotes variance. Combining this result with the signal of the target image (ignoring the  pedestal), E$[I^2]$, this gave the SNR as:
\begin{align}
    \text{SNR} = \sqrt{\frac{\text{E}[I^2]}{ \text{Var} \left[ P' \right]}}.
\end{align}

\begin{table*}
\centering
\caption{Key parameters associated with the six Maxipix-based X-ray ghost projections, in Fig.~3(f) and 4(a-e) from the main paper. In the left column, we use the following abbreviations: ``No.''=``number'', ``Exp.''=``exposure time'', ``SD''=``standard deviation'', and ``SNR''=``signal-to-noise ratio''.}
\label{table:Mask/Image GP}
\begin{tabular}{|l|c|c|c|c|c|c|} 
\hline
{\em Image} & Fig. 3(f) & Fig. 4(a) & Fig. 4(b) & Fig. 4(c) & Fig. 4(d) & Fig. 4(e) \\ 
\hline
{\em No.~Pixels} & 1250 & 100  & 676 & 1296 & 1296 & 2112 \\ 
\hline
\begin{tabular}[c]{@{}l@{}}{\em No.~Available}\\{\em Masks}\end{tabular} & 17\,280 & 484  & 5\,760 & 35\,574 & 35\,574 & 29\,645 \\ 
\hline
\begin{tabular}[c]{@{}l@{}}{\em No.~Selected}\\{\em Masks}\end{tabular}  & 820 & 92 & 265 & 490 & 462 & 388 \\ 
\hline
{\em Mean Exp.~(ms)} & 20.1 & 33.0 & 23.0 & 18.5 & 24.0 & 26.2 \\ 
\hline
{\em SD Exp.~(ms)} & 17.3  &27.4 & 20.6 &  17.4 & 23.3 & 25.2 \\ 
\hline
{\em Max Exp.~(ms)} & 118  & 167  & 125 & 136  & 136 & 136 \\ 
\hline
{\em Min Exp.~(ms)} & 1.01 & 1.19 & 1.07 & 0.07 & 0.18 & 0.27 \\ 
\hline
\begin{tabular}[c]{@{}l@{}}{\em Experimental}\\{\em SNR}\end{tabular} & 3.01 & 5.04 & 2.44 & 1.56 & 1.71 & 1.16 \\
\hline
\end{tabular}
\end{table*}

\begin{table*}
\begin{center}
\caption{Breakdown of which masks were employed for each of the Maxipix ghost projections, in Figs.~3(f) and 4(a-e) from the main paper.  Here, the column titles are abbreviated as follows: {\em F40} -- random-fractal mask with 40 $\mu$m feature size, {\em F20} -- random-fractal mask with 20 $\mu$m feature size, {\em R40} -- binary random mask with 40 $\mu$m feature size,  {\em G20} -- binarized Gaussian-smoothed-noise mask with 20 $\mu$m feature size, {\em L20} -- binarized Lorentzian-smoothed-noise mask with 20 $\mu$m feature size, {\em FRT40} -- finite-Radon-transform based Legendre mask with 40 $\mu$m feature size and $p = 127$, $N'$ -- number of selected masks.}
\label{table:WhichMasksWereUsedForWhichGPs}
\begin{tabular}{ |c|c|c| c | c | c | c| c | } 
\hline
{\em Target} & {\em F40} & {\em F20} & {\em R40} & {\em G20} & {\em L20} & {\em FRT40} & $N'$ \\ 
\hline
Fig.~3(f) & 259 & 492 & - & - & 69 & - & 820 \\ 
\hline
Fig.~4(a) & 92  & - & - & - & - & - & 92 \\ 
\hline
Fig.~4(b) & - & - & - & - & 265 & - & 265 \\ 
\hline
Fig.~4(c) & - & 164 & 140 & 35 & 101 & 50 & 490 \\ 
\hline
Fig.~4(d) & - & 132 & 135 & 57 & 75 & 64 & 463 \\ 
\hline
Fig.~4(e) & - & 149 & - & 86 & 128 & 25 & 388 \\ 
\hline
\end{tabular}
\end{center}
\end{table*}

\section*{Supplement 1, Section 8: Media files}

Here we describe the ghost-projection videos that accompany our paper.  In each of these videos, the left panel shows the individual Maxipix exposures that make up a given ghost projection, with the cumulative digital-detector exposure given in the right frame.  
\begin{itemize}
\item Visualization 1 corresponds to the ``GP'' image shown in Fig.~3 of the main paper, together with the more detailed version in Fig.~\ref{fig:Four Build up GPs==SUPP} and Fig.~\ref{fig:six digital GPs==SUPP}(c) of this Supplement. 
\item Visualization 2 corresponds to Fig.~4(a) in the main paper (positive-contrast dot), together with the more detailed version in Fig.~\ref{fig:six digital GPs==SUPP}(a) of this Supplement. 
\item Visualization 3 corresponds to Fig.~4(b) in the main paper (2 positive-contrast and 2 negative-contrast squares), together with the more detailed version in Fig.~\ref{fig:six digital GPs==SUPP}(b) of this Supplement. 
\item Visualization 4 corresponds to Fig.~4(c) in the main paper (negative-contrast smiley face), together with the more detailed version in Fig.~\ref{fig:six digital GPs==SUPP}(d) of this Supplement. 
\item Visualization 5 corresponds to Fig.~4(d) in the main paper (positive-contrast smiley face), together with the more detailed version in Fig.~\ref{fig:six digital GPs==SUPP}(e) of this Supplement. 
\item Visualization 6 corresponds to Fig.~4(e) in the main paper (negative-contrast segment of the ESRF logo), together with the more detailed version in Fig.~\ref{fig:six digital GPs==SUPP}(f) of this Supplement. 
\end{itemize}

\bibliography{UniversalXrayMask==arXiv-version}

\end{document}